\documentclass[aps,prl,preprint]{revtex4-1}
\usepackage{graphicx}
\usepackage{amsmath}
\usepackage{siunitx}
\usepackage{caption}
\usepackage{bm}
\usepackage{textcomp, gensymb}
\usepackage{hyperref}
\hypersetup{colorlinks = true,linkcolor=blue,
     filecolor=blue,
     urlcolor=blue,
     citecolor = blue, pdfauthor=author}
\begin{document}


\title{Gear-based 3D-printed Micromachines Actuated by Optical Tweezers}


\author{Alaa M. Ali}
\affiliation{Université Marie et Louis Pasteur, SUPMICROTECH, CNRS, Institut FEMTO-ST, F-25000 Besançon, France}
\email{alaa.ali@femto-st.fr}
\author{Gwenn Ulliac}
\affiliation{Université Marie et Louis Pasteur, SUPMICROTECH, CNRS, Institut FEMTO-ST, F-25000 Besançon, France}
\author{Edison Gerena}
\affiliation{MovaLife Microrobotics, Paris, France}
\author{Abdenbi Mohand-Ousaid}
\affiliation{Université Marie et Louis Pasteur, SUPMICROTECH, CNRS, Institut FEMTO-ST, F-25000 Besançon, France}
\author{Sinan Haliyo}
\affiliation{Sorbonne Université, CNRS, Institut des Systèmes Intelligents et de Robotique (ISIR), Paris, France}
\author{Aude Bolopion}
\affiliation{Université Marie et Louis Pasteur, SUPMICROTECH, CNRS, Institut FEMTO-ST, F-25000 Besançon, France}
\author{Muamer Kadic}
\affiliation{Université Marie et Louis Pasteur, SUPMICROTECH, CNRS, Institut FEMTO-ST, F-25000 Besançon, France}



\begin{abstract}
The miniaturization of mechanical mechanisms is crucial to enable the development of compact, high-performance micromachines. However, the downscaling actuation of conventional gears and micromotors has remained limited by the inherent challenges of implementing mechanical/electrical powering. Here, we present the design, fabrication, and characterization of an optomechanical, gear-driven micromachine realized through two-photon polymerization 3D printing. The actuation is achieved using optical tweezers.  The device integrates a microgear transmission system with an optically actuated part, enabling light-controlled micromachines. When illuminated by a highly focused laser source, the first gear generates rotational torque within the gear assembly, converting optical energy into directional mechanical work that can be transmitted to the coupled gear. We demonstrate the fabrication of micromachines using two-photon polymerization (2PP) laser writing, enabling the fabrication of spur gear trains and bevel gears that can produce out-of-plane rotations, which is not achievable with traditional micromachining fabrication techniques. The micromachines are composed of a single gear or a train of two or three gears without any unwanted adhesion between the components, leading to functioning systems. Experimentally, the fabricated micromachines were actuated using optical tweezers, demonstrating continuous gear rotation, effective motion transmission in gear trains, out-of-plane rotations, and the ability to amplify velocity or torque. Optical-tweezer actuation broadens the potential applications of these micromachines, particularly in biomedical and lab-on-a-chip systems, where precise, minimally invasive control at the microscale is essential.
\end{abstract}

\maketitle

\section{Introduction}
Microgears have evolved as fundamental components for microscale power transmission and motion conversion. Their ability to transmit motion, amplify torque, and couple multiple degrees of freedom has positioned them as important components of next-generation microrobotic and lab-on-chip systems. Microgears can be actuated using different techniques. For example, electrostatic-driven microgear design has been studied ~\cite{arai2024power} to characterize the power transmission efficiency in microgear trains. Moreover, electrostatics can be combined with optical force in optoelectronic tweezers to drive polymeric microgears ~\cite{zhang2021reconfigurable}. However, using electrostatic force requires electrical contacts and wires, which limits their use. Therefore, other contactless actuation methods were proposed; for instance, magnetically responsive microgears fabricated from composite polymer with magnetic nanoparticles~\cite{kavre2014magnetic, peng2024magnetic} enable non-contact operation and gear-to-gear coupling in viscous media. Furthermore, acoustic excitation further expands the design space, as surface-acoustic-wave driven micro-rotors can act as drag-based gears for contactless fluidic pumping~\cite{ding2012saw}. Additionally, optical strategies have demonstrated versatile actuation.  Recently, Wang \textit{et al.}~\cite{wang2025metamachines} reported microscopic geared micromachines in which metasurface structures are used to redirect the optical force to generate torque and actuate gear trains with sub-micrometer precision, bringing optical actuation and photonic design into mechanical functionality. An optically driven microgear transmission system can be used as pumps as proposed by Wu \textit{et al.}~\cite{wu2025opticalgear}, where a set of microrotors is assembled and coupled via optical, hydrodynamic interactions to translate microbeads. Together, these works illustrate that microgears serve as the mechanical backbone of micromachines. However, the actuation of microgear trains using optical tweezers has not yet been demonstrated, as achieving stable optical trapping and torque transfer on complex multi-component structures remains experimentally challenging, particularly because their fabrication requires careful optimization to ensure proper alignment of components while avoiding adhesion or fusion.
\\

Optical tweezers employ a tightly focused laser beam to trap and manipulate microscopic objects by transferring photon momentum \cite{alaee2016optically,alaee2018opticalPRB}. The optical field exerts two primary forces: a gradient force, which pulls dielectric particles toward the region of highest light intensity near the beam focus, and a scattering force, which pushes them along the direction of light propagation due to radiation pressure. The balance between these forces creates a stable three-dimensional optical trap that enables controlled manipulation of the objects \cite{ashkin1986observation}. Although optically driven micromachines were recently reported based on a non-focused laser source directed to a metasurface that generates rotational motion \cite{wang2025metamachines}, the integration of optical tweezers for micromechanical actuation offers several advantages. First, optical tweezers provide flexibility in system control, enabling, for instance, the selection of the driving and driven gears within a gear train without the need to predefine this configuration during fabrication. In addition, the ability to generate multiple optical traps allows the simultaneous manipulation of several components within the same system. Optical tweezers are also commonly used in biological applications~\cite{Edison}, as the thermal and photonic damage can be minimized by confining the light of the tightly focused laser   to localized micro-spots directed toward a micromachine, rather than exposing large areas that may lead to harmful laser interactions with biological cells. Accordingly, micromachines actuated by optical tweezers can be effectively implemented in lab-on-chip applications, enabling functionalities such as cell manipulations and flow generations together with metamaterials and fluid and vibration controlled devices \cite{Piest2024,Dudek2025}. Furthermore, optical tweezers mainly actuate transparent polymeric structures, which can be fabricated using three-dimensional printing by two-photon polymerization (2PP), providing design flexibility and enabling the integration of 3D-structures and components at different vertical levels without the planar constraints imposed by conventional photolithography. Therefore, this approach also opens opportunities for the construction of more complex microelectromechanical systems (MEMS) with multiple degrees of freedom \cite{pagliano2022micro3D}, and micromachines able to give motion in-plane or out-of-plane directions.
 
However, the usage of optical tweezers in microrobots and micromachines manipulation also presents two major challenges. First, the magnitude of the optical force is relatively small-typically in the range of piconewtons \cite{andrew2022amplification, williams2002opticalforce,xin2020opticalforce} which limits the achievable mechanical power and torque output. Second, conventional optical tweezers systems inherently provide a limited number of degrees of freedom, as they generally constrain particle motion to planar translation or in-plane rotation within the focal plane \cite{OT_theory&practice}. Nevertheless, these restrictions can be overcome by specialized optical tweezers setups that can allow the three-dimensional control of the  traps \cite{EdisonIEEE}, or if the optically manipulated object is designed to induce additional degrees of freedom, such as out-of-plane rotation or three-dimensional motion \cite{Alaa2025Chiral,outofplaneAntoine,outofplaneMaria}.
Therefore, gear-based micromachines actuated by optical tweezers can be advantageous as they can mechanically amplify torque, convert translational motion into rotational motion (or vice versa), and even transform in-plane motion into out-of-plane motion. Thus, by coupling multiple gears with optical tweezers, it becomes possible to enhance the mechanical functionality and expand the degrees of freedom of optically driven microrobotic systems.

\begin{figure}[!h] 
    \centering
     \includegraphics[width=17cm]{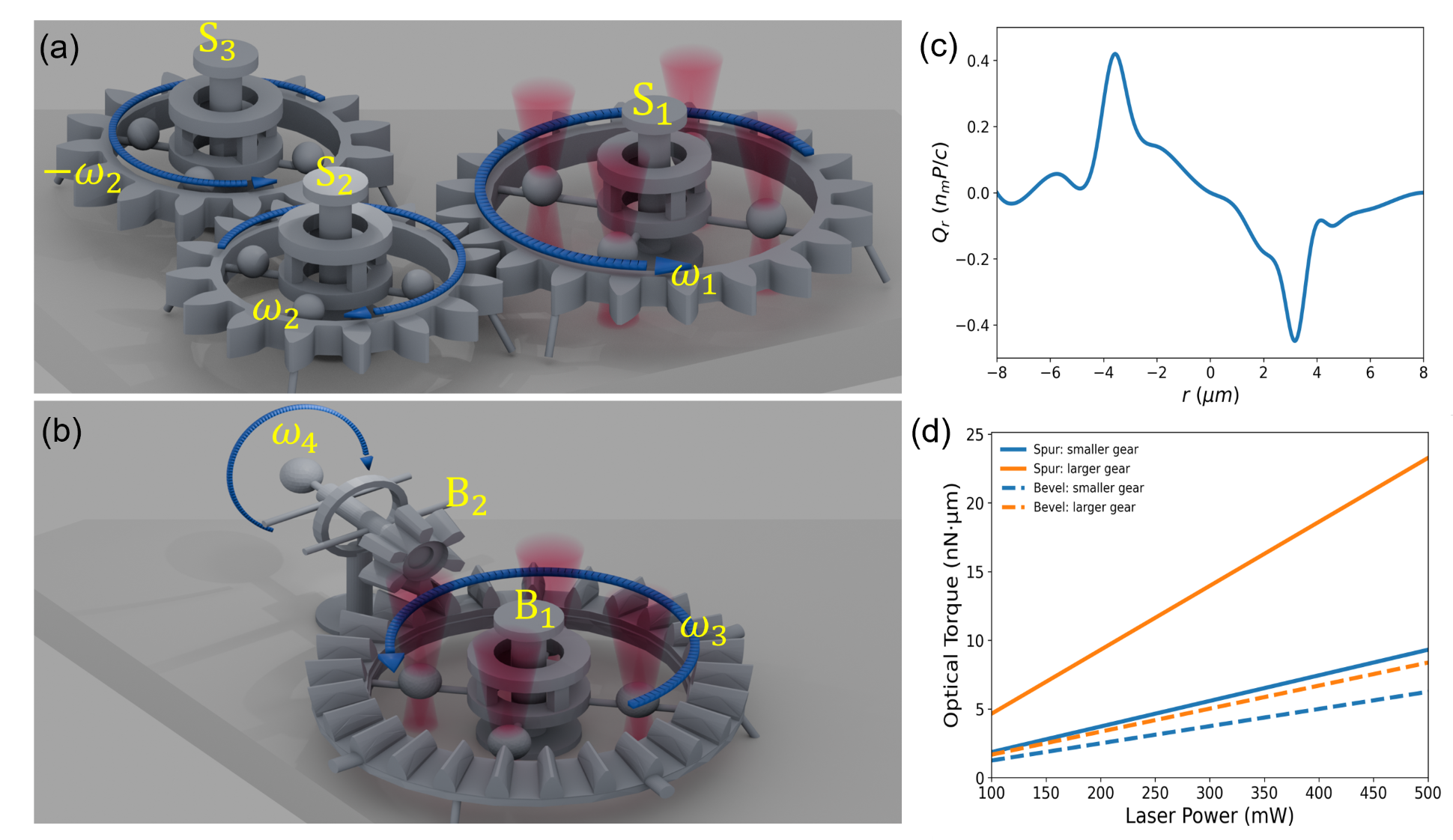}
    \caption{Principle of the gear-based micromachines actuated by optical tweezers for controlled rotational motion transmission. (a) A train of spur gears to give in-plane rotation. A large spur gear ($S_1$) is actuated by multiplexed optical traps that move in a controlled circular motion of velocity $\omega_1$, and the motion is transferred to the small gears ($S_2$ and $S_3$), which rotate with an amplified velocity of magnitude $ \omega_2$. (b) A system of bevel gears to give out-of-plane rotations; the larger bevel horizontal gear ($B_1$) is rotated in-plane using the optical tweezers, and accordingly the smaller one above it ($B_2$) is rotated in out-of-plane direction with higher velocity $\omega_4$. (c) The radial optical trapping efficiency (the normalised optical force) on the spherical optical handles of $5 \mu $m diameter is shown versus the position of the trap (calculated using the T-matrix method). (d) The calculated optical torque for the two-gear system for different laser power, for the spur gears, the calculations are done assuming one of the smaller gears is actuated by optical traps and the bigger gear is driven ($S_1$); while in the bevel gears, the larger gear ($B_1$) is the driving one, and accordingly $B_2$ is driven.}
    \label{fig1}
\end{figure}

In this paper, we introduce gear-based micromachines actuated by optical tweezers for motion transmission. We present two types of micromachines. Firstly, spur gears that can rotate in-plane and be included in trains of gears with different diameters for motion transmission, and torque or velocity amplification (fig. \ref{fig1}(a)). Secondly, bevel gears to transmit the in-plane rotation into out-of-plane rotation with amplified velocity (fig. \ref{fig1}(b)). Each planar gear is equipped with four spherical optical handles that serve as trapping sites. These handles are individually trapped using time-multiplexed focused laser beams, which are then translated in a circular trajectory to induce controlled rotational motion of the gear; so it acts as a driver gear and accordingly, the other(s) are driven. 

 Microrobots actuated by optical tweezers are most commonly fabricated by three-dimensional printing using the 2PP technique, as it enables the creation of fully three-dimensional microstructures with sub-micron precision using polymeric transparent dielectric materials suitable for optical trapping. However, fabricating multi-component systems using 2PP remains challenging \cite{tan2022single}, since closely spaced parts are highly prone to unwanted adhesion or fusion during the printing or development process. This often results in a fixed or partially immobilized micromachine rather than a freely movable assembly \cite{outofplaneMaria}. Several strategies have been proposed to mitigate these effects. For instance, the introduction of sacrificial support structures has been shown to allow the fabrication of movable parts within a single printing step \cite{SacrificialStructures}, while careful post-development washing can remove unpolymerized material and release freestanding components \cite{WashAwayResin}. Additionally, supercritical CO$_2$ drying is often employed to reduce capillary stiction and prevent collapse of delicate features. \cite{SupercriticalDrying}.

Therefore, in this work, we present a gear-based micromachine designed for controlled motion transmission. We perform optical force and torque calculations. Furthermore, we demonstrate the fabrication of fully functional, multi-component micromachines with non-fused and mechanically independent parts, achieved through the optimization of two-photon polymerization parameters and structure design. Finally, we experimentally validate their controlled motion and torque transmission using an optical tweezers setup, confirming the feasibility of actuating multi-component, gear-based micromachines by optical tweezers. This work opens the door to new functionalities and applications in optically driven micromechanical systems. To the best of our knowledge, this is the first reported demonstration of a gear-train micromachine actuated by optical tweezers, and a bevel gear that can give continuous out-of-plane rotation in the microscale.

\section{Results and Discussions}
\subsection*{Design of the microgear and optical force calculations}\label{sec2}
The proposed micromachines presented in this work consist of two or three interconnected gears, where one gear is in-plane optically actuated (Figure 1), and the others rotate accordingly through mechanical coupling. Two types of systems are presented; the first one is based on a spur-gear train to transmit in-plane rotations (fig.~\ref{fig1}(a)), and the second one is based on bevel gears to convert in-plane rotation into out-of-plane rotation (fig.~\ref{fig1}(b)). In each system, one gear acts as the driving gear, which is actuated using the rotation of optical traps, while the other gears are driven accordingly.

Here, we show torque and velocity amplification as a result of gear transmission. Typically, when the smaller gear serves as the driving gear, it transfers motion to the larger one, which rotates with an amplified torque. Conversely, when the larger gear acts as the driving gear, the smaller one rotates with a higher angular velocity, effectively amplifying the rotational speed. This behaviour follows the fundamental gear ratio principle, where the ratio of angular velocities is inversely proportional to the ratio of gear diameters $\left( \frac{\omega_1}{\omega_2} = \frac{D_2}{D_1} \right)$, while the torques are inversely related $\left( \frac{\tau_1}{\tau_2} = \frac{D_1}{D_2} \right)$. Hence, mechanical power is conserved $\left( \tau_1 \omega_1 = \tau_2 \omega_2 \right)$, and by selecting appropriate gear dimensions, the trade-off between torque and speed can be tailored at the microscale. The concept of optical torque amplification is particularly important, as the torque exerted by light is inherently limited to low values and therefore requires enhancement mechanisms~\cite{xu2024gradient, articleturbine}. Moreover, the generation of out-of-plane rotation from planar motion is also demonstrated, which is essential since the motion of optical traps is mostly restricted to planar movement.

In the spur-gear system, two different gear sizes are used; when the smaller gear is the driver, the torque transmitted to the larger gear is amplified, whereas when the larger gear drives, the angular velocity of the smaller gear is amplified. As shown in fig.~\ref{fig1}(a), each gear is composed of a rotating part and a stator, which consists of an anchored rod at the centre. The larger gear has a diameter of $50~\mu$m, while the smaller one has a diameter of $37.5~\mu$m. The velocity amplification factor is equal to the ratio between the two gear radii, $r_\text{big}/r_\text{small}=4/3\approx1.33$, whereas the torque amplification factor corresponds to the ratio between the larger gear radius and the distance of the optical handle from the centre of the smaller gear, $r_\text{big}/r_\text{optical handle}\approx2.5$.

In the bevel-gear system, the planar microgear must act as the driving gear, since the movement of optical traps is limited to planar motion. Consequently, the out-of-plane rotation of the upper gear is only allowed when the lower planar gear is rotated. In this case, the velocity ratio between the two gears is determined by the ratio between the number of teeth of each gear, which is $25/6\approx4.167$. However, the torque ratio is $\approx1.34$ as it depends on the diameter of the optical handles positions where the optical traps are applied to give the rotation and the diameter of the smaller gear. \\

The optical force responsible for actuating the gear handles is evaluated using the Transition Matrix (T-matrix) method implemented in the \emph{Optical Tweezers Toolbox} (OTT)~\cite{OTT}. The simulation considers a $5\,\mathrm{\mu m}$-diameter dielectric sphere, representing the optical handle ($n=1.5$), trapped by a tightly focused Gaussian beam propagating in water ($n_m=1.33$). OTT is used to calculate the optical trapping efficiency parameter $Q$, which is a dimensionless quantity representing the momentum transfer efficiency between the incident photons and the trapped particle. As such, $Q$ is commonly regarded as a normalized optical force and is used to compare optical forces independently of the laser power and the surrounding medium. The resulting optical force is therefore given by
\begin{equation}
F = Q \frac{n_m P}{c},
\end{equation}
where $P$ is the laser power at the focus, $n_m$ is the refractive index of the surrounding medium, and $c$ is the speed of light in vacuum. Figure~\ref{fig1}(c) shows the calculated variation of the normalized efficiency $Q$ as a function of the sphere displacement from the focal point, with a maximum value of $Q_\text{max}=0.42$.

For an experimentally estimated laser power of $P=100$--$500~\text{mW}$, the optical force reaches $F=185~\text{pN}$ at $P=100~\text{mW}$. In the spur-gear system, when the smaller gear is the driver, the corresponding torque acting on the smaller gear is $1.85~\text{nN·µm}$, calculated based on the diameter at the optical handle position. The optical torque is then amplified by the larger gear to $4.5~\text{nN·µm}$, calculated based on the full diameter of the larger gear, resulting in a torque amplification factor of $2.4$. For the bevel gear configuration, a torque transmission ratio of 1.34 is obtained. For the bevel gear system, an input torque of approximately 1.68 nN·µm applied to the larger driving bevel gear results in a transmitted torque of about 1.25 nN·µm on the smaller driven gear, as in this case, the velocity is amplified. Increasing the optical power linearly enhances both the trapping force and the generated torque, as summarized in fig.~\ref{fig1}(d). This optical torque is resisted by drag force and friction; therefore, the resulting rotational velocity depends on their net effect.

\subsection{Fabrication Process Optimizations}
\begin{figure*}[!h] 
    \centering
     \includegraphics[width=19cm]{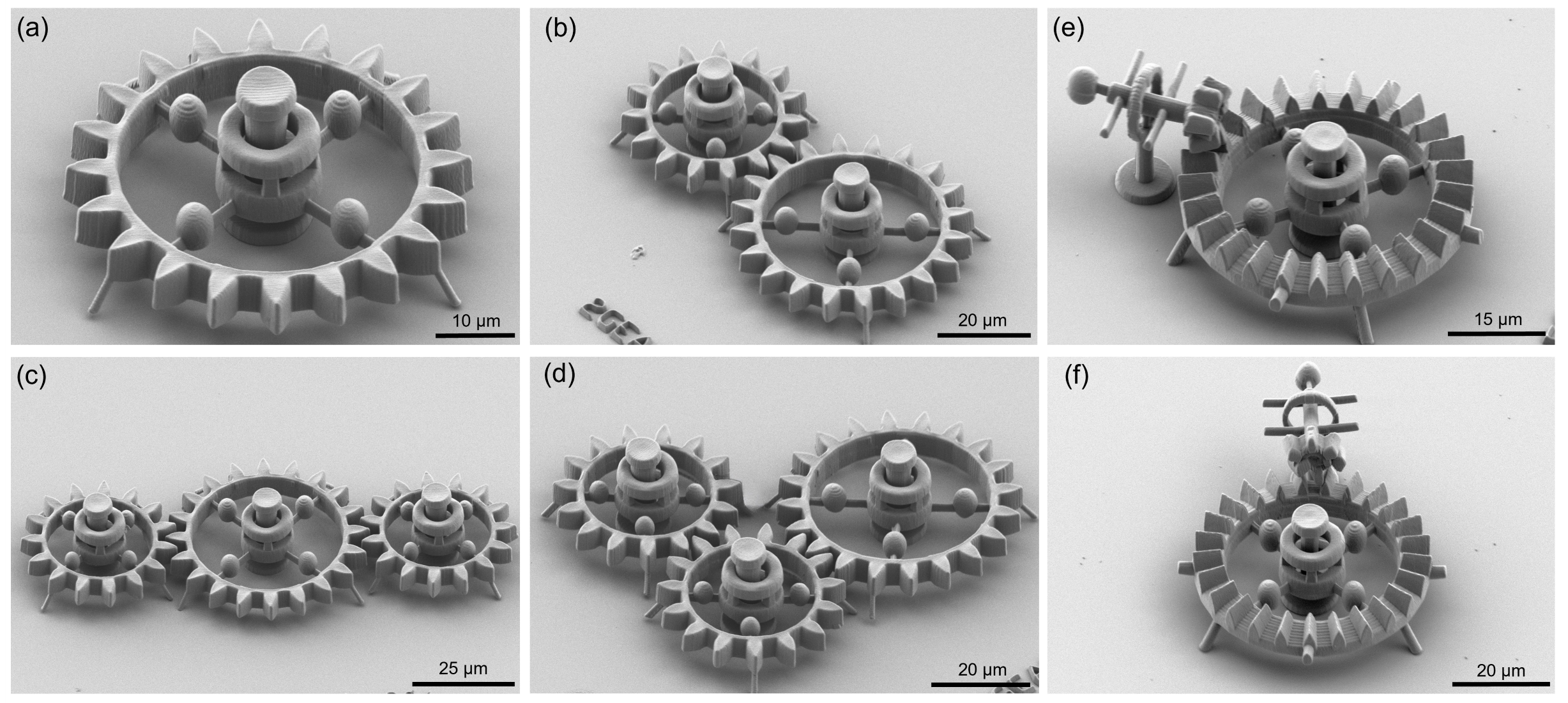}
    \caption{SEM images of the fabricated micromachines by 3D-printing using 2PP.  (a) a single microgear with a central stator and a surrounding rotor gear. The stator is a rod anchored to the substrate with a wide base and a cap. The rotor is a gear with two rings surrounding the stator, connected by four rods between them. The gear also has four spherical optical handles to direct the optical traps on, in addition to four pillars connecting the gears to the substrate for support during the fabrication and to be mechanically broken after to get a free-rotating gear. (b) a micromachine consisting of a train of two microgears. (c-d) Two designs of a three-microgear train. (e-f) Two different orientations of a micromachine composed of a horizontal bevel gear coupled to a smaller gear positioned above it, enabling out-of-plane rotational motion. All the images were taken while the substrate is tilted by an angle $45\deg$, shown from different orientations to show that the micromachines are efficiently fabricated without adhesion or fusion between their components.}
    \label{SEM}
\end{figure*}
The micromachines were fabricated using a three-dimensional direct laser writing system based on the 2PP technique. The Scanning Electron Microscope (SEM) images presented in fig. \ref{SEM} show a spur microgears system: a single gear, a train of two gears, and two designs of a three-gear train. It also shows a system of bevel microgears for out-of-plane rotations. The single gear consists of a stator and a rotor. The stator is a rod anchored to the substrate with a wide base to ensure good adhesion to the substrate and a cap to maintain the gear in place during operation. The rotor is a gear with two rings over each others surrounding the stator connected by four rods between them. The inclusion of two rings, rather than a single one, reduces tilting of the gear during rotation, as the dual-ring design constrains the maximum allowable angular deviation compared to a single ring. This structural feature guarantees improved alignment and stability when the gear is integrated into a multi-gear train. The design of two rings also allows getting rid of any unpolymerised resin between the rotor and stator during the development, as if they are replaced by one tall ring, the unpolymerized resin may be trapped between the stator and the rotor, leading to adhesion between them. Each gear has four spherical optical handles to direct the optical traps, in addition to four pillars connecting the gears to the substrates to support each gear during the fabrication process, but they are mechanically detached from the substrate after the complete fabrication to allow free rotation of the gears around the stator.\\

The 2PP method employs a tightly focused femtosecond laser to trigger localized polymerization within a photosensitive polymer through the simultaneous absorption of two photons. Because the nonlinear absorption occurs only at the focal volume, polymerization is confined to a well-defined voxel, allowing structures to be written with submicrometer resolution (fig.S1 (a)). By scanning the laser focus throughout the resist in three dimensions, complex 3D microstructures are built sequentially in a layer-by-layer manner, making this approach suitable for the realization of intricate microsystems. However, to fabricate a multi-component micromachine actuated by optical tweezers, there are two main challenges: i) The micromachine stators should be attached to a thin transparent coverslip suitable for high-resolution microscopy and optical tweezers experiments. ii) The multi-component structures are prone to adhesion between its components due to the fusion between the components, surface tension of liquid while drying, or residues of unpolymerized resin that get trapped between the components leading to their adhesion. \\

To address the first challenge, the microgears were fabricated using the oil-immersion configuration. In general, two main optical configurations can be used in 2PP: the dip-in (fig. S1(b)) and oil-immersion modes (fig. S1(c)). In the dip-in configuration, the objective lens is directly immersed in the liquid photoresist, allowing the laser to focus several micrometers below the resin surface without an intermediate substrate. This setup is advantageous for printing tall, free-standing 3D structures, as it eliminates spherical aberrations arising from refractive index mismatches. However, the polymer used in this configuration requires fused silica substrates -to ensure refractive index contrast- which are mostly available in high thicknesses, if cost-effective substrates are needed. As a result, the printed structures are not easily compatible with optical tweezers systems, which typically require thin, transparent coverslips for trapping laser transmission. Conversely, in the oil-immersion configuration, a thin borosilicate glass coverslip (typically 170 µm thick) is used as a substrate, over which a uniform photoresist layer is added. This approach is suitable for fabricating structures intended for optical tweezers experiments, as it provides a flat and optically transparent base compatible with high-NA objectives that need thin substrates, allowing the focal point of the trapping laser beam to be over the substrate and not inside it. The immersion oil, whose refractive index is matched to both the glass and the photoresist, minimizes optical aberrations and ensures precise focusing throughout the polymerization volume. The resulting microstructures adhere to the coverslip, preventing unwanted displacement during optical trapping, while maintaining optical clarity for efficient transmission of the trapping beam.\\

Therefore, the oil-immersion configuration was selected as the most suitable approach to fabricate the microgears. Nevertheless, this approach may cause incomplete fabrication of the upper portions of the gears (fig. S3). This is attributed to the fact that in this configuration, the laser must pass through the resin and polymerizes the structure layer by layer; as the writing height increases, optical aberrations and focal shift caused by refractive‐index mismatch and cover-slip thickness can degrade the focusing conditions and thus reduce polymerisation efficiency \cite{Stichel2016, Burmeister2015}. To mitigate these effects, we employed a combination of two remedial strategies: (i) implementing a defocusing method (offsetting the nominal focal plane) to compensate for the axial focal drift while also conducting power slope so that as we go to the higher layers the laser power becomes higher (for more information see supplementary materials), and (ii) optimizing the overall height of the structure so that the critical top features remain within the optimal focal region of the objective, but at the same time, maintaining sufficient axial spacing between adjacent components to prevent adhesion or fusion, and ensuring adequate joint height to preserve mechanical stability and prevent tilting during rotation. These adjustments enabled reliable fabrication of the full gear geometry.

For the second challenge, the design of the microgears had to be carefully optimized to achieve the most suitable geometry and laser parameters that prevent adhesion between components and enable successful rotation. This optimization involves selecting appropriate clearances within the joint region (between the stator and rotor) and tooth-to-tooth spacing between adjacent gears; large enough to prevent fusion or sticking (fig. S5), yet sufficiently close to allow effective mechanical coupling and torque transfer. 

Furthermore, the structures must be mechanically supported to avoid any floating of layers during the fabrication. Floating refers to the movement of the polymerized layers while the complete printing process is not terminated yet. This happens if the printed layer is so thin or if it is not well supported by the layers beneath it, connected to the substrate. Floating or deformed layers can merge unintentionally with nearby components, resulting in undesired adhesion between the different components of the micromachine. Therefore, -as mentioned before- the gears were mechanically supported during fabrication using pillars connected to the substrate to prevent tilting or floating of partially polymerized layers. The laser-writing parameters also play a crucial role: excessively small slicing and hatching distances improve print resolution for single, compact structures but can be detrimental for multi-component assemblies, as they increase fabrication time and raise the likelihood of layer deformation or floating before complete polymerization. Therefore, slicing and hatching distances were optimized for the fabrication of our micromachine Figure S2. Finally, the laser power must be precisely tuned to balance polymerization efficiency and feature fidelity, avoiding both over-exposure (which can cause fusion between parts) (fig. S2) and under-exposure (which results in incomplete or fragile structures). Moreover, printing of each component should be done in a separate step, so instead of printing the whole structure in a single step, the stators should be printed first, then each gear (fig. S4). 
The optimization of all the discussed parameters is essential for the successful fabrication of fully functioning micromachines. Conversely, if any of these parameters is not properly addressed, even a single point of adhesion within the micromachine can be sufficient to render it non-functional.

Finally, to allow free rotation of the microgears, the pillars are mechanically broken using a micromanipulator under a light microscope. This step serves as a preliminary test for the fabrication efficiency. When a gear is pushed by the micromanipulator, it rotates, leading to the detachment of the pillars and the driving of the next coupled gears, if the fabrication is successful and there is no adhesion between its components. However, if there is adhesion between the components, the microgears are not rotated, and pushing by the micromanipulator may lead to breaking or complete separation of the whole micromachine from the substrate.
\begin{figure}[!h] 
     \includegraphics[width=19cm]{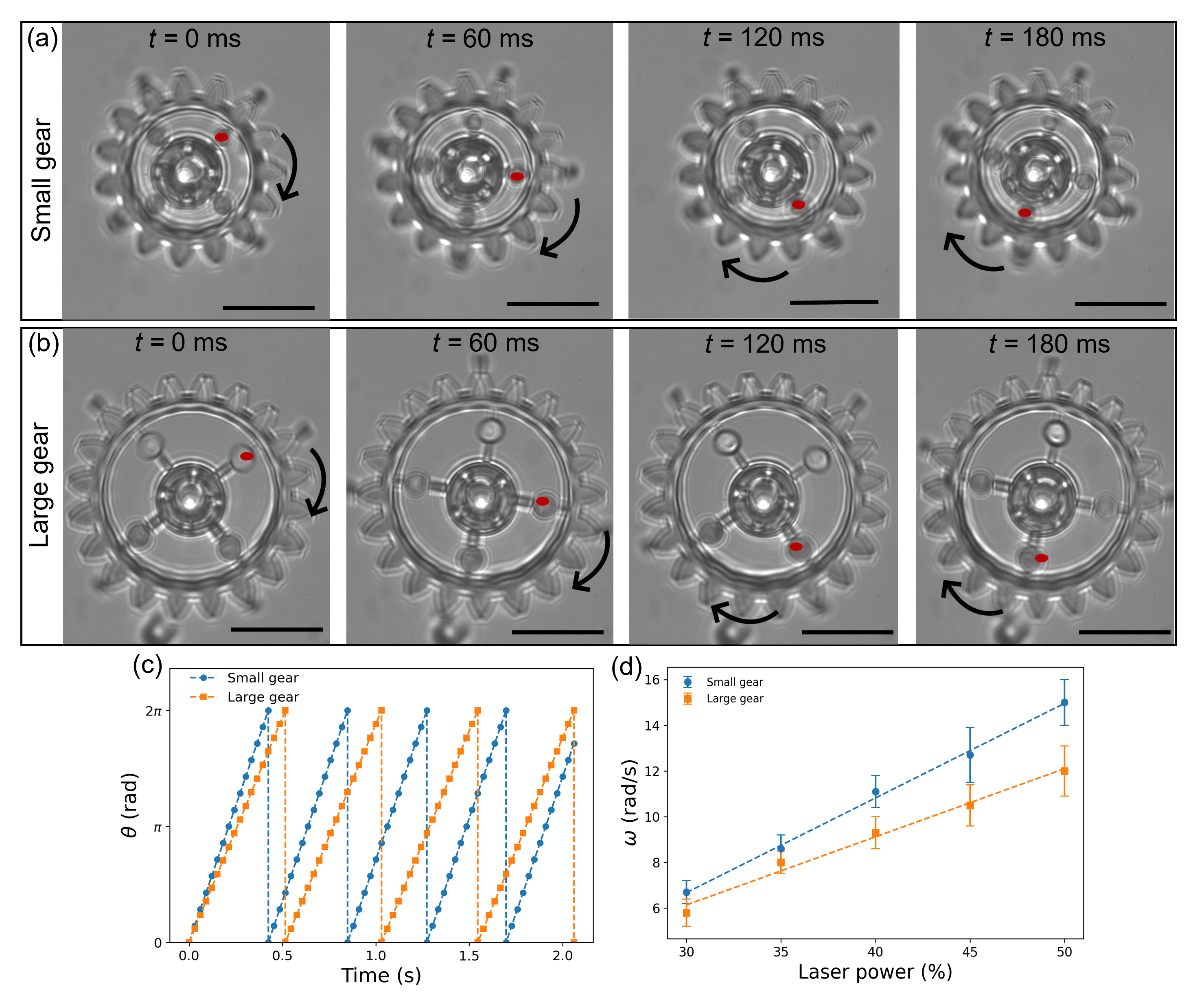}
    \caption{Rotation of single gears. (a,b) Time-lapse optical microscopy snapshots of the (a) small gear and (b) large gear rotating at their maximum angular velocity, obtained at a laser power of $P=50\%$. The rotation is followed by tracking a red marker placed on one of the optical handles of the gear, which serves as a visual reference for the angular displacement over time. The gears rotate in a clockwise direction. Consecutive frames are separated by approximately 60~ms. The rotational motion is induced by steering multiplexed optical traps. Scale bars: $20\mu m$. (c) Illustration of angular displacement as a function of time for the small and large microgears at maximum used laser power, showing continuous rotation. (d) maximum obtainable angular velocity $\omega$ as a function of laser power for the small and large gears. Each data point corresponds to the mean value obtained from three measurements, while the error bars represent the standard deviation. Dashed lines represent a linear fit. Some of the measured velocities here are shown in video S1.}
    \label{onegear}
\end{figure}
\section{Optical Tweezers Experiments}
In our experiment, we show the rotation of the gears with the optical tweezers system to apply multiplexed optical traps on the optical handles positions as shown before in fig. \ref{fig1}, and allow circular motion with controlled speed. Time multiplexing means that a single optical trap is rapidly scanned between different positions at a frequency much higher than the mechanical response of the system, resulting in a simultaneous trapping of multiple handles.The direction and amplitude of motion are computed on the basis of the vector defined by the center of the four traps and the position of each trap, along with a scaling factor that sets the velocity. The laser is deflected from one trap to another at a constant frequency, causing the traps to move in discrete steps; higher velocities correspond to larger step sizes. As the four optical handles are connected within the structure, the gear remains in continuous motion \cite{Edison}.

The optical tweezers system is shown in fig. S6. The optical traps can hold the optical handles and control their motion. The speed of the rotation depends on the optical trapping force and the resisting forces, such as the drag and friction. As shown before in Figure 1, the optical torque is dependent on the laser power; therefore, when the laser power increases, the maximum rotational velocity at which the gear is able to follow the optical trap's motion also increases. In our experiments, we used a laser setup where the maximum output laser power is $10$ W; however, the laser power is measured before the objective lens and around $80\%$ are lost even before passing through the objective and reaching the sample therefore, the estimated power that reaches the sample is $10 \%$ of the output value. Moreover, the maximum output power percentage used in the optical tweezers system is $50\%$ to avoid overheating of the manipulated samples. In our experiments, we worked with power $30-50\%$; so this corresponds to a power range of $300-500 m$W. 
\\
\subsection{Spur microgears for in-plane rotations}
In fig. \ref{onegear}, we firstly show the rotation of a single microgear rotating alone around its stator and not coupled to any other gear around it. In panel (a), screenshots are taken from video S1 for the rotation of the small gear at its maximum velocity, which is $15.7$ rad/s. For the larger gear, the maximum velocity is $12.6$ rad/s. The difference in the maximum obtainable velocity for both gears can be attributed to the higher drag in the case of the larger gear. In fig. \ref{onegear}(d), it is also shown that the dependence of the maximum obtainable rotational velocity on the laser power is linear with the power. When trying to increase the velocity of rotation of the optical traps higher than the values indicated in the graph for each laser power, the microgear fails to follow their rotations; therefore, misalignments occur between the optical traps and optical handles, leading to unstable motion of the microgear around the stator.

\begin{figure*}[!h] 
    \centering
          \includegraphics[width=19cm]{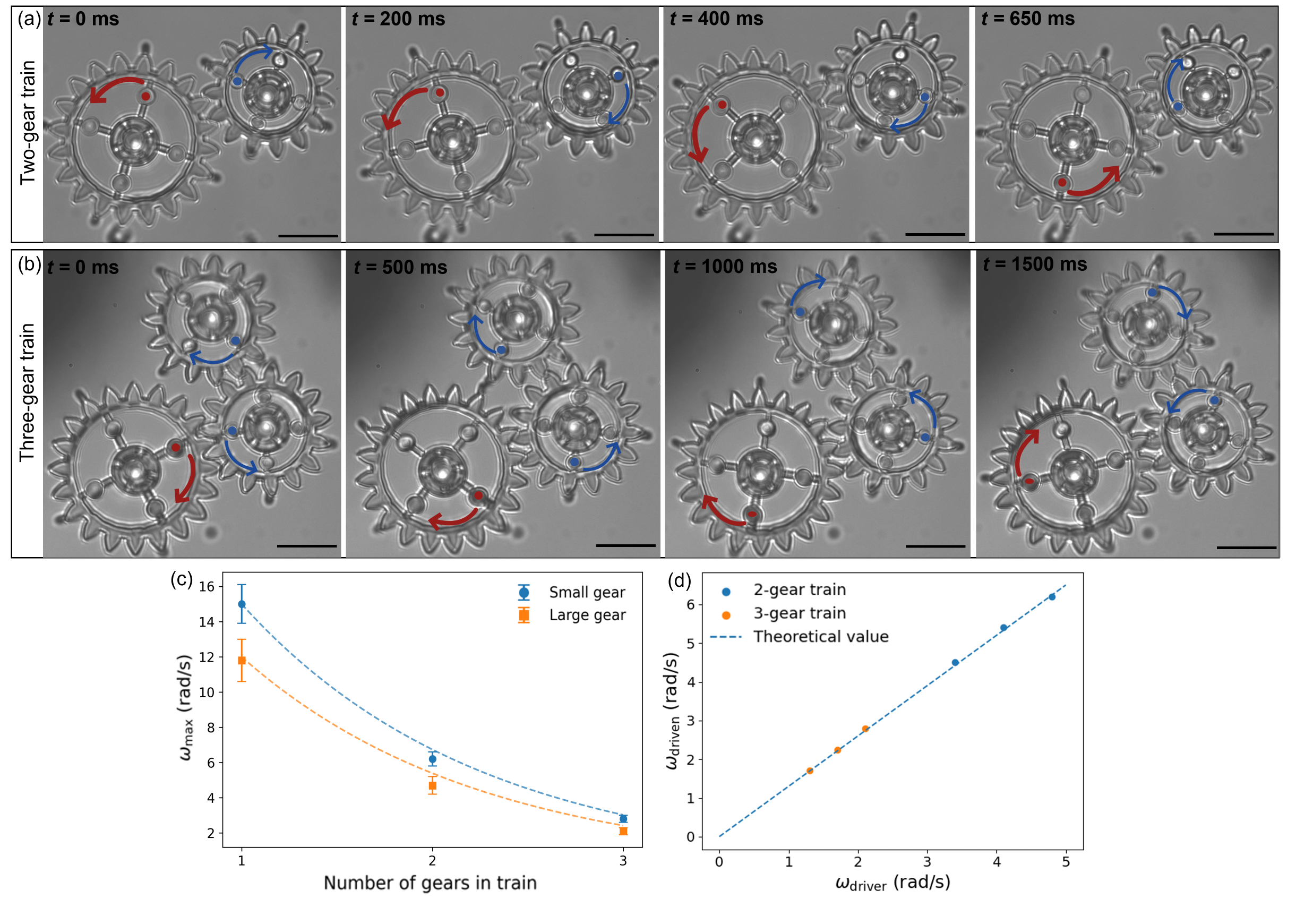}
    \caption{Rotation of optically actuated microgear trains. (a) Time-lapse optical microscopy images of a two-gear train showing counter-rotating motion of the large (driver) and small (driven) microgears under multiplexed optical trapping at maximum velocity which is also shown in in part 4 of Video S2. (b) A three-microgear train actuated at maximum velocity illustrating sequential transfer of rotation through the meshed gears. Screenshots are taken from Video S3 part 4. The direction of rotation is indicated by curved arrows, while coloured dots mark selected optical handles used to track angular displacement over time. The driver gears are indicated by red dots and arrows, while the driven gear is indicated by blue ones. Here, we show the anti-clockwise rotation of the driver microgear in the two-gear train and its clockwise rotation in the three-gear train to indicate the freedom in choice of the rotation directions.
The screenshots show half of a complete rotation of the driver gear. Scale bars: 20~\textmu m. (c) Maximum angular velocity $\omega_{\max}$ of individual gears as a function of the number of gears in the train. The decrease in $\omega_{\max}$ with increasing train length reflects the cumulative mechanical load, giving an exponential drop. Error bars denote the standard deviation. (d) The angular velocities of the driven and as a function of the driver angular velocity for two- and three-gear trains. The dashed line indicates the theoretical value. The measured velocities are taken from three rotations, giving the same values. The close agreement between experimental data and the theoretical value over the investigated speed range evidences stable meshing and efficient motion transfer with negligible slip.
}
    \label{train}
\end{figure*}
 Secondly, we show the performance of a micromachine composed of a gear train of two or three microgears actuated by optical tweezers. When microgears are mechanically coupled to form a gear train, one of the gears is chosen to be the driving gear, where the multiplexed optical traps are directed to its optical handles to control its rotation, while the other(s) are driven gears affected by the driving gear's motion.
 
 Efficient displacement transformation is observed experimentally in both two-gear and three-gear trains (fig. \ref{train}(a,b)); however, the maximum rotational velocity at maximum laser power starts to decrease approximately exponentially with increasing the number of microgears in the train (fig. \ref{train}(c)). This behavior can be attributed to the cumulative mechanical losses along the gear train, due to increased friction at the meshing interfaces. Therefore, the maximum velocities for a two-gear train are $6.3$ and $4.8$ rad/s for the small gear and large gear, respectively, and in the three-gear train, the maximum velocities decrease to $3.1$ and $2.24$ rad/s.\\
 In fig. \ref{train}, we show representative video frames where the larger microgear is used as the driving gear, and the smaller ones are driven. Additional actuation capabilities for a two-gear micromachine are shown in video S2; which highlights the freedom in the choice of the direction of the rotation, in the choice of the driver and the driven gear and also in the choice of the velocity of rotation up to the maximum velocity. The same applies for the three-gear micromachine (video S3), we fabricated two designs (fig. \ref{SEM}(c)(d)) and both show good meshing between the gears and efficient transformation of the displacement for the two driven gears and the velocity amplification.\\ 
 
 Due to fabrication limits, there is a distance of $2 \mu$m between the rotor and stator, which may lead to pushing of the driven microgears away from the driving one; resulting in some misalignments. To overcome this and improve meshing stability, in our experiments, an additional fixed optical trap is also placed near the centre of the driven gears. This biases their position more towards the driver microgear, and prevents its movement away, leading to better meshing and preventing any misalignments. Therefore, as shown in fig. \ref{train}d, the experimental velocities of the driven microgears are very close to what is calculated using the ratio between the diameters of the larger microgear and the smaller one.

\subsection{Bevel microgears for Out-of-plane rotation}
\begin{figure*}[!h] 
    \centering
     \includegraphics[width=19cm]{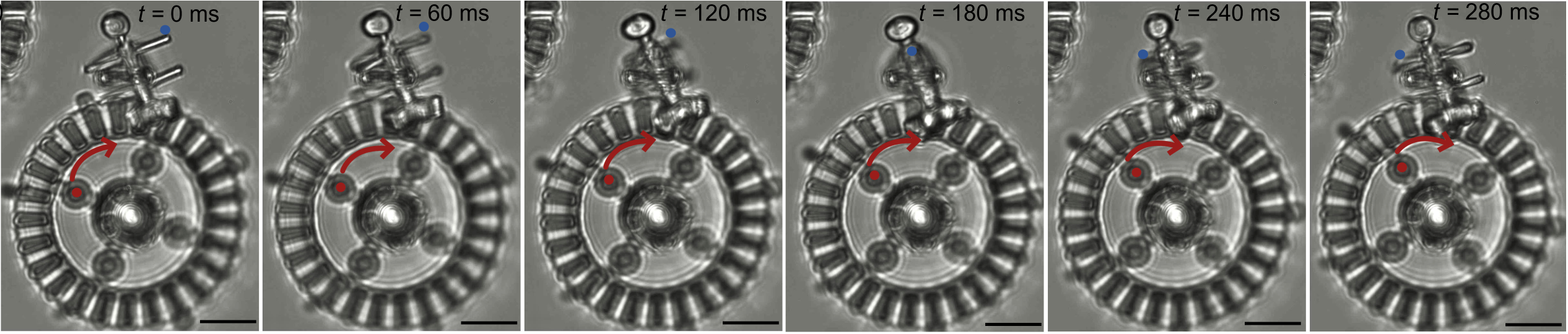}
    \caption{Time-lapse optical microscopy images showing the out-of-plane rotation of the driven bevel gear actuated by the in-plane rotation of the driver gear. The blue marker indicates a reference point on the driven gear arm, while red arrows highlight the rotation direction. The screenshots are taken from part 3 of video S4. Scale bar is $10\mu$m}
    \label{outofplane}
\end{figure*}

\begin{figure*}[!h] 
    \centering
     \includegraphics[width=9cm]{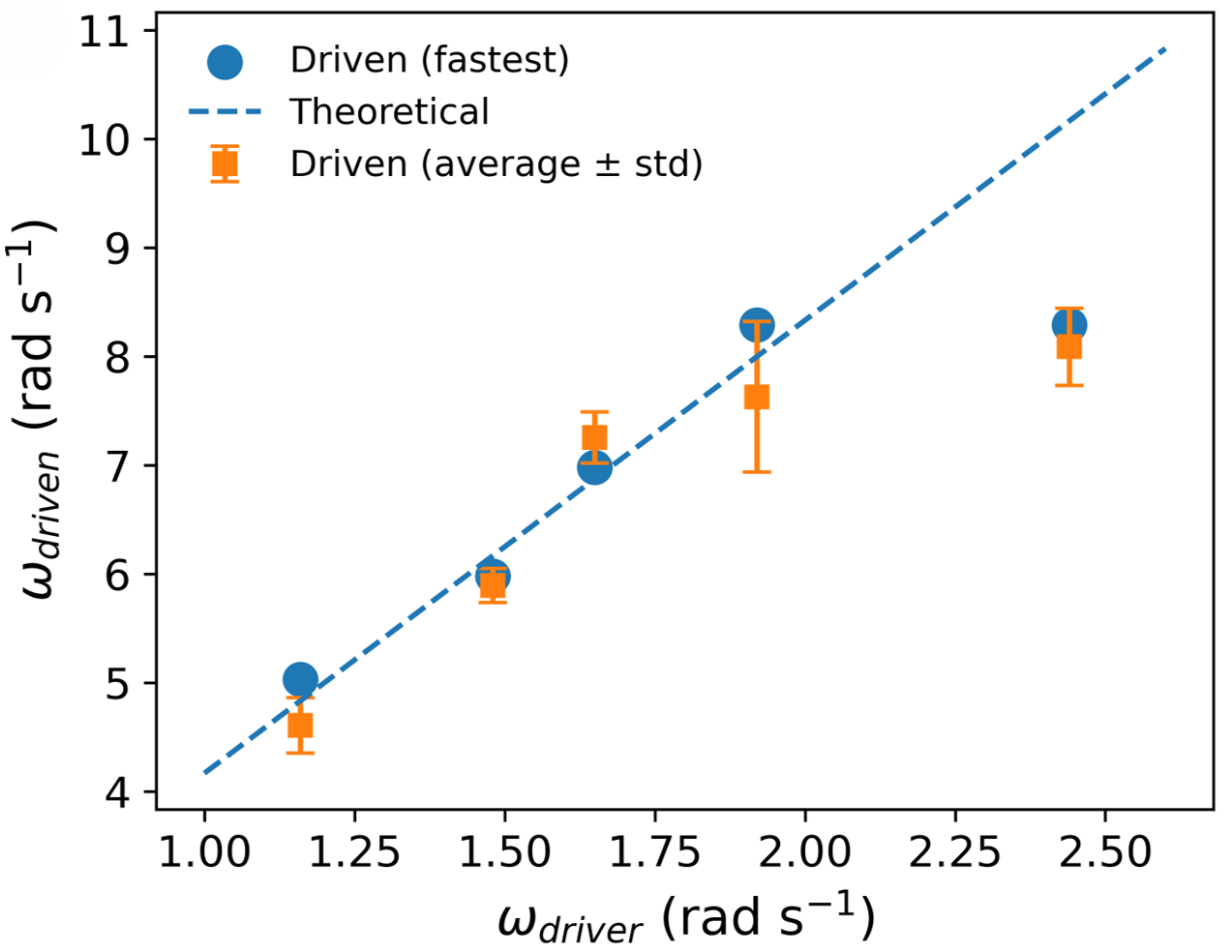}
    \caption{Angular velocity of the driven gear, $\omega_{\mathrm{driven}}$, as a function of the driver angular velocity, $\omega_{\mathrm{driver}}$. Circles correspond to the fastest measured rotations, whereas squares represent averages over three consecutive turns. The dashed line indicates the theoretical transmission predicted from the number of teeth of the bevel gears. Deviations from the theoretical trend are attributed to intermittent misalignment during out-of-plane engagement. Some of the velocities here are shown in video S4.}
    \label{outofplane2}
\end{figure*}

Achieving out-of-plane rotation in microscale systems remains challenging, particularly when relying on conventional planar micromachining techniques. In this work, the use of three-dimensional printing enables the fabrication of a bevel-gear-based micromachine composed of two microgears. The lower gear is optically actuated in-plane using rotating optical traps and acts as the driving gear, while a second gear positioned above it is mechanically driven to rotate out-of-plane.

The upper gear is mechanically constrained by a surrounding ring aligned with its rotation axis, forming a joint that limits undesired motion. In addition, an optical trap is applied to its optical handle to further stabilize the structure and suppress unintended in-plane or out-of-plane displacement. fig. \ref{outofplane} presents video frames illustrating the in-plane rotation of the lower gear and the corresponding out-of-plane rotation of the upper driven gear, which is visualized using arms oriented perpendicular to the rotation axis.

The relationship between the angular velocity of the driving gear (at maximum laser power) and the resulting out-of-plane angular velocity of the driven gear is shown in fig. \ref{outofplane2}. Under optimal meshing conditions, the driven gear reaches angular velocities close to the theoretical value predicted from the gear tooth ratio. These cases correspond to the data points labeled “fastest” in fig. \ref{outofplane2} and represent the maximum velocity achieved for a given driving speed, occurring when gear meshing is most efficient.

However, deviations from the theoretical behavior are observed due to occasional misalignment between the gears. At higher driving speeds, lateral displacement or slight tilting of the upper gear reduces the meshing efficiency, leading to a lower average angular velocity of the driven gear. As a result, the driven gear velocity begins to fall below the theoretical prediction for driving speeds exceeding $2.5$ rad/s, and at higher speeds the motion becomes unstable, with incomplete rotations observed. Consequently, the maximum stable angular velocity of the driven gear is observed to be $8.29$ rad/s. The rotation of the bevel-gear micromachine is shown in Video S4 at the maximum velocity and at other lower velocities.

In contrast to in-plane spur gear configurations -where meshing remains efficient and the performance is primarily limited by optical force, viscous drag, and friction- the bevel gear configuration is mainly constrained by alignment accuracy between the gears. This limitation can be attributed to fabrication constraints: the ring forming the joint around the upper gear is relatively wide, as it also incorporates a supporting pillar required during fabrication and later removed before experiments. Future design optimizations targeting tighter joint tolerances are expected to further improve alignment and enable higher stable out-of-plane rotational velocities.
\section{Conclusion}

In this work, we demonstrated a fully 3D-printed, gear-based micromachine that is actuated contactlessly using optical tweezers. By combining two-photon polymerization with an optically addressable gear design, we achieved reliable fabrication of single gears and multi-gear trains with mechanically independent components (i.e., without adhesion or fusion), enabling continuous rotation and efficient motion transmission. Spur-gear trains enabled controllable in-plane rotation with torque/velocity transformation through gear ratios, while a bevel-gear architecture converted in-plane actuation into continuous out-of-plane rotation -an important capability presented for the first time in the microscale to our knowledge- as it is difficult to realize with conventional planar micromachining. Experimental optical-tweezer actuation confirmed stable meshing, predictable speed ratios, and mechanical amplification, with performance mainly limited by viscous drag, friction, and alignment tolerances in the bevel configuration. Overall, these results establish optical tweezers as a versatile, minimally invasive power source for complex microscale transmissions and open new opportunities for light-controlled micromachines in lab-on-a-chip and biomedical environments where precise, reconfigurable actuation is required.

\section{Experimental Section}
\subsection{Numerical simulations}
Transition matrix (\emph{T-matrix})
formalism was used to calculate the optical trapping efficiency implemented in the MATLAB Optical Tweezers Toolbox (OTT)\cite{Nieminen_OT_Theory}.

This method exploits the linearity of Maxwell’s equations to describe the light
scattering process as a linear operator that relates the incident and scattered
electromagnetic fields. It provides an efficient framework for calculating
optical forces and torques, particularly for particles possessing geometrical or
material symmetry, as the same precomputed T-matrix can be reused for multiple
particle positions and orientations without recalculating the full
electromagnetic field
\cite{Nieminen2011}.

Within this formalism, the scattering process is expressed as
\begin{equation}
\mathbf{E}_{\text{sca}} = \mathbf{T}\,\mathbf{E}_{\text{inc}},
\label{eq:t_operator}
\end{equation}
where $\mathbf{E}_{\text{inc}}$ and $\mathbf{E}_{\text{sca}}$ denote the incident
and scattered electromagnetic fields, respectively, and $\mathbf{T}$ is the
transition matrix of the particle. The elements of the T-matrix fully encode the
particle morphology, material properties, and their interaction with the incident
field.

The T-matrix approach is particularly suitable for the present work, as it
enables accurate evaluation of optical trapping efficiency in the intermediate
(Mie) regime where the optically trapped object size is comparable to the trapping laser wavelength. Moreover, it is suitable for the symmetric structures so it was employed to calculate the optical forces and
torques acting on the spherical optical handles of the fabricated micromachines
under the experimental trapping conditions.

\subsection{Fabrication}
The micromachines were fabricated using a commercial 3D laser lithography system (Photonic Professional GT+, Nanoscribe GmbH) based on the 2PP technique. The negative photoresist IP-L (Nanoscribe GmbH) was selected as the printing material, suitable for fabrication in the oil-immersion configuration on borosilicate glass substrates (D263) with a thickness of $(170 \pm 10)\mathrm{\mu}$m. The micromachines were printed layer by layer along their longitudinal axis, using slicing and hatching distances of $0.1 \mu$m. A drop of photoresist was placed on the substrate and polymerized with a femtosecond laser ($\lambda = 780,\mathrm{nm}$) through a 63× oil-immersion objective. The laser power was set to 30\%, with a galvanometric scan speed of $5000 \mu$m/s for the entire structure except the first several layers of the base of the stator, which was printed with laser power of 40\% and the same scanning speed to ensure strong adhesion to the substrate. Each component was laser-written in a separate step, so the stators were printed first, then each gear, to avoid any adhesion between them.  
Following fabrication, the samples were developed for 12 minutes in propylene glycol methyl ether acetate (PGMEA) to remove unpolymerized resin, then rinsed twice in isopropyl alcohol (IPA) for 3 minutes each to eliminate any remaining developer. Finally, the micromachines were dried using critical point drying (Autosamdri 931, Tousimis) to prevent liquid surface tension that may lead to adhesion between the micromachine components.

\subsection{Pillars detachment from the substrate}
To detach the pillars connecting the microgears to the substrate to allow the subsequent rotation with the optical tweezers, a micromanipulator was used. A platform centered around a commercial optical microscope (OPTIKA) was employed. The setup incorporated a three-axis micromanipulator system (MP-$285$, Sutter Instrument) equipped with a micro-rod fabricated from borosilicate glass (B$150$-$110$-$10$). The micro-pipette tip was formed using a micropipette puller (Sutter Instruments, Model P-$1000$ Flaming/Brown) to achieve an approximate diameter of $2~\mathrm{\mu m}$. The pipette was mounted on the micromanipulator, allowing precise control of its movement to gently push individual microgears, allowing their rotation, which induces the detachment of the pillars from the substrate, and the driving of the coupled gears in the gear trains.

\subsection{SEM Imaging}
The fabricated microromachines were characterized using a ThermoFisher ApreoS SEM operated in Optiplan mode. To ensure high-quality imaging, the samples were sputter-coated with a thin chromium layer ($\approx$ 5~nm) using a Leica ACE 600 $600$ sputter coater. The acceleration voltage was kept below 3 kV to avoid structural damage. The substrate was tilted to angle of $45^\circ$ to get a side view of the micromachines. The SEM images were obtained from structures that were nominally identical to those employed in the optical experiments; however, they were not the exact same samples, as the metallic coating required for high-resolution SEM imaging would otherwise alter their optical behavior.

\subsection{Optical manipulation}

The optical tweezer platform, illustrated in Supplementary fig.S6, combines an optical trapping system, a precision motion stage, and a control interface. The optical system is built around a custom inverted microscope equipped with an oil-immersion objective (Olympus UPlanFLN $40\times$, NA~$1.3$) and a near-infrared laser source ($\lambda = 1070~\mathrm{nm}$). LED illumination provides visual feedback, captured by a high-speed CMOS camera (Basler, $659 \times 494~\mathrm{px}$). 

The setup enables two complementary actuation modes. A $3$D nanopositioning stage (PI~P-$562.3$CD) mounted on a $2$D micro-stage (PI~M-$126$.CGX) provides a large workspace and fine positional control, while dynamic beam steering is achieved through a galvanometric mirror pair (GVS$002$, Thorlabs) and a deformable mirror (PTT$111$~DM, Iris~AO) for rapid, time-shared multi-trap manipulation. The system allows precise sample control via the controller interface, coordinating micro- and nano-stage motion while maintaining fixed optical traps -thereby combining active and passive actuation for enhanced manipulation accuracy. Moreover, the setup provides a speed mode that allows defining the motion of the optical traps along a circular path while controlling their speed; therefore, this mode was used to actuate the microgears by directing the optical traps to the optical handles, then moving them along a circular path. ~\cite{Edison}. The micromachines were actuated inside a medium consisting of deionized water containing $1\%$ Pluronic, which prevented adhesion between micromachine components.

\section*{Data availability}
The datasets used and/or analyzed during the current study 
 are available from the corresponding author on reasonable request.

\section*{Author contributions statement}
A.M.A. and M.K. conceived the study and the design.
A.M.A has designed, performed numerical computations, conducted experimental work, and written the first draft. G.U. has fabricated the samples. A.M.A. and E.G. have conducted the experimental validations. E.G., A.M.-O., S.H., A.B., and M.K. have contributed to the overall project supervision, discussions, and paper writing.

\section*{Competing interests}
The authors declare no competing interests.
\section*{Materials and Correspondence}
Correspondence and requests for materials should be addressed to A.M.A. (Email:alaa.ali@femto-st.fr).
\section*{Additional information}
To be included as a link.

\medskip
\textbf{Supporting Information} \par Supporting Information is available from the Wiley Online Library or from the author.

\medskip
\textbf{Acknowledgements} \par 
The authors acknowledge the support of the ANR OPTOBOTs project (ANR-21-CE33-0003) and the ANR PNanoBot (ANR-21-CE33-0015). The work was supported by the French RENATECH network and its FEMTO-ST technological facility MIMENTO, by the French research infrastructure ROBOTEX (TIRREX ANR-21-ESRE-0015) and its FEMTO-ST technological facility CMNR and by the French-Swiss SMYLE Network. 

\medskip

\bibliographystyle{MSP}


\end{document}